# Radiation resistance of fine-grained ceramics $Y_{2.5}Nd_{0.5}Al_5O_{12}$ under Xe-ions irradiation

Alekseeva L.S.[a], Nokhrin A.V.[a, (*)], Yunin P.A.[b], Nazarov A.A.[a,b], Orlova A.I.[a], Skuratov V.A.[c], Issatov A.T.[c,d,e], Kovylin R.S.[a,f], Murashov A.A.[a], Boldin M.S.[a], Voronin A.V.[a], Chuvil'deev V.N.[a], Zotov D.A.[a]

[a] Lobachevsky State University of Nizhny Novgorod, Nizhny Novgorod, 603022 Russia

[b] The Institute for Physics of Microstructures RAS, Nizhny Novgorod, 603950 Russia

[c] Flerov Laboratory of Nuclear Reactions, Joint Institute for Nuclear Research (FLNR JINR), Dubna, Moscow oblast, 141980 Russia

[d] Gumilyov Eurasian National University, Nur-Sultan, 010000 Kazakhstan

[e] The Institute of Nuclear Physics, Almaty, 050032 Kazakhstan

[f] G.A. Razuvaev Institute of Organometallic Chemistry RAS, Nizhny Novgorod, 603950 Russia

albina.orlova@gmail.com, golovkina_lyudmila@mail.ru

**Abstract:** Oxide $Y_{2.5}Nd_{0.5}Al_5O_{12}$ with garnet structure was synthesized in the powder and ceramics forms. Fine-grained ceramics with a relative density of ~99% were obtained by the Spark Plasma Sintering method. The radiation resistance of ceramics was studied under irradiation with swift Xe-ions (E = 146 MeV). Full amorphization of the samples was observed under irradiation with a fluence of $1×10^{13}$ cm$^{-2}$. The calculated value of the critical fluence was $(6.5 ± 0.1) × 10^{12}$ cm$^{–2}$, which corresponded to 0.03 dpa. The microhardness $H_V$ of the surface layer decreased with increasing ion fluence.

**Keywords:** Garnet ceramics; Spark plasma sintering; Radiation resistance; hardness.

---

(*) Corresponding authors (nokhrin@nifti.unn.ru)

# 1. Introduction

The creation of a closed nuclear fuel cycle is an urgent task of modern nuclear power engineering. The solution of this problem will allow to reduce the volume of generated radioactive waste and to recycle existing reserves of plutonium and minor actinides (MA): Np, Am, Cm. One of the ways to reduce the environmental hazard of high-level radioactive waste (HLW) is to isolate MA for its transmutation into stable or short-lived isotopes during irradiation in thermal or fast neutron reactors [1-4]. Within the framework of this task, it is important to develop inert matrix fuels (IMF) - uranium-free fuel with an inert matrix containing MA [5-11]. The main requirements for IMF are high melting and/or decomposition temperatures, high thermal conductivity, absence of phase transformations at operating temperatures, resistance to radiation exposure, etc. [6, 12, 13].

Crystalline compounds with a garnet structure are promising for creating a new generation of IMFs [5, 14-16]. Iso- and heterovalent substitutions of the main atoms of the crystal lattice with lanthanides and actinides are possible in the garnet structure [17]. The crystal lattice of garnet can contain high concentrations of lanthanides and actinides, in particular up to 18 wt.% uranium [18, 19]. In the Upper Chegem caldera (Northern Caucasus, Russia), the natural uranium garnet elbrusite containing up to 27 wt.% $UO_3$ was found [20]. Compounds with a garnet structure are highly stable in aqueous solutions and have high radiation resistance [21-23]. The use of mineral-like compounds for IMF, including those based on garnet, will further provide direct geological disposal of MA without significant recycling.

One of the main requirements for new IMF materials is their high radiation resistance to the effects of neutrons, helium ions formed during the α-decay of actinides and to the effects of fission fragments. To date, the structure and radiation properties of garnet-based IMFs have been quite well studied after irradiation with neutrons, α-particles (including internal irradiation) and swift heavy ions (SHI) of low energies (1-2 MeV) [15, 19, 24, 25]. Since the impact of fission fragments is characterized by a high level of specific ionization energy losses, it can be modeled using high-energy heavy ion beams at accelerators.

In [15], radiation resistance of $Y_3Al_5O_{12}$ (YAG) garnet was studied by irradiation with neutrons (E > 0.1 MeV) to a fluence of $1.7 \times 10^{26}$ m$^{-2}$ at 815 K. The change in the volume of YAG after irradiation was less than 1%. Transmission electron microscopy (TEM) results revealed no differences between the samples before and after irradiation. In [18], the susceptibility of the garnet structure to self-irradiation from α-decay of $^{244}$Cm and resistance to irradiation by Kr$^{2+}$ ions with an energy of 1 MeV were investigated. Garnets of various compositions were studied: $(Y_{2.43}Nd_{0.57})(Al_{4.43}Si_{0.44})O_{12}$, $(Ca_{1.64}Ce_{0.41}Nd_{0.42}La_{0.18}Pr_{0.18}Sm_{0.14}Gd_{0.04})Zr_{1.27}Fe_{3.71}O_{12}$, $(Ca_{1.09}Gd_{1.23}Ce_{0.43})Sn_{1.16}Fe_{3.84}O_{12}$, and garnet $Y_{2.89}Cm_{0.1}Pu_{0.01}Al_5O_{12}$ doped with 3 wt.% $^{244}$Cm. It was established that the $Y_{2.89}Cm_{0.1}Pu_{0.01}Al_5O_{12}$ garnet became amorphous at an irradiation dose corresponding to 0.4 dpa; amorphization upon irradiation with an ion beam was observed in the range of 0.29–0.55 dpa. In [24], $Y_{3-x}Nd_xFe_5O_{12}$ garnet (x = 0.1, 1.8) was irradiated with α-particles with an energy of 2 MeV with fluences from $1 \times 10^{14}$ to $1 \times 10^{17}$ ions/cm$^2$ at room temperature. The formation of an amorphous layer with a thickness of 84 nm was observed, although the grain shape and element distribution remained uniform after irradiation. The neodymium content did not affect the radiation resistance of the garnet structure. In [26], $Y_3Fe_5O_{12}$ (YIG) garnet was irradiated with Cu (50 MeV), S (50 MeV), and Kr (235 MeV) ions. Upon Kr irradiation, the YIG phase gradually amorphized due to increased track overlap. Partial recrystallization of amorphous tracks occurred upon irradiation with S and Cu, which led to the formation of grains sizes around 10 nm. It is concluded that these processes are initiated by losses of electronic energy. In general, the authors of [26] note that the radiation stability of the garnet structure is not very sensitive to changes in its composition.

The purpose of this work is to study the radiation resistance of ceramics based on $Y_{2.5}Nd_{0.5}Al_5O_{12}$ (YAG:Nd) oxide with a garnet structure under irradiation with swift heavy Xe ions. Nd was used as a chemical and structural analogue of Am/Cm. The ceramics were produced by Spark Plasma Sintering (SPS), which is a variation of the high-speed hot-pressing method [27]. The advantage of this method is the ability to reduce the sintering temperature and holding time, which is especially important while working with radioactive materials. This allows the formation of a fine-grained microstructure in ceramics with a high relative density [28]. It was previously shown that garnet-based ceramics obtained by the SPS method have high hydrolytic stability [6, 29, 30]. There have been no studies of the radiation resistance of fine-grained YAG:Nd ceramics.

## 2. Materials and Methods

Garnet powder with the composition $Y_{2.5}Nd_{0.5}Al_5O_{12}$ (YAG:Nd) was obtained by coprecipitation method according to the procedure described in [31].

Ceramics were produced by the SPS method on the Dr. Sinter model-625 setup. The powders were placed in a graphite mold with an internal diameter of 12 mm and heated by passing millisecond pulses of high-power electric current (up to 3 kA). Sintering was carried out in vacuum, at T = 1390 °C and P = 70 MPa. The heating rate to a temperature of 600 °C was 100 °C/min, to a sintering temperature of 1390 °C - 50 °C/min. There was no isothermal holding at the sintering temperature. Temperature was measured using a Chino IR-AH pyrometer focused on the surface of a graphite mold. The accuracy of temperature determination was ±10 °C, the accuracy of maintaining the applied pressure was 1 MPa. Ceramics sintering diagrams are shown in **Figure 1**.

Sintered samples were annealed in air at 750 °C (2 h) to remove graphite residues from the surface, and then were mechanically polished.

The density of the sintered samples ($\rho$) was measured by hydrostatic weighing in distilled water using a Sartorius CPA scales. The average accuracy of determination of $\rho$ was 0.01 g/cm$^3$. The theoretical density ($\rho_{th}$) of the synthesized compounds was calculated based on the analysis of the results of X-ray studies.

XRD experiments on phase analysis of ceramic samples were performed on a Bruker D8 Discover X-ray diffractometer in symmetric Bragg-Brentano geometry. An X-ray tube and CuK$\alpha$ radiation were used. The radius of the goniometer is 30 cm. All experiments were performed under the same conditions. The size of the primary beam in the equatorial plane was limited by a 0.6 mm wide slit, and in the axial plane – 12 mm when taking a diffraction pattern. Diffraction patterns were recorded by $\theta/2\theta$ scanning in the angular range from 10° to 90° along the 2$\theta$ angle. The angle step was 0.05°. A LynxEYE linear position-sensitive detector with 192 independent imaging channels and an angular aperture of 2° over 2$\theta$ was used. The accumulation time at the point was 2 s, which is equivalent to an effective accumulation time of about 40 s at the point due to the use of a position-sensitive detector. For qualitative phase composition, the PDF-2 database was used.

The microstructure of ceramics was studied using scanning electron microscopes (SEM) Tescan Vega 2 and Regulus SU8100.

Radiation tests of ceramic samples were carried out on the cyclotron IC-100 of the Laboratory of Nuclear Reactions of the Joint Institute for Nuclear Research (Dubna, Russia). Ceramic samples were irradiated with Xe ions with an energy of 146 MeV with fluences from $1\times10^{12}$ to $1\times10^{13}$ ions/cm$^2$. The average ion flux was $2\times10^9$ cm$^{-2}$s$^{-1}$. Irradiation was carried out at room temperature. The surface temperature of the ceramic sample did not exceed 30 °C during irradiation. The accuracy of measuring ion flux and fluence was 15%.

## 3. Results and Discussion

Four identical samples of YAG:Nd ceramics with a diameter of 12 mm and a height of 3 mm were produced using the SPS method. The samples had no major defects (pores, cracks, chips). The density of the ceramic samples was 4.67-4.68 g/cm$^3$ (98.4-98.8% of the theoretical density of the YAG:Nd phase). The differences in the density of the samples did not exceed three measurement errors (less than 0.03 g/cm$^3$).

Ceramics have a homogeneous fine-grained microstructure (**Figure 2**); the average grain size of garnet (dark gray areas) is 5-10 μm, the particle size of the perovskite impurity phase (light gray areas) is 1-2 μm. Also, there is residual porosity in the structure - black areas with a size of ~0.5-1 μm.

XRD studies of ceramics (**Figure 3**) show that each of the YAG:Nd samples contains a phase isostructural to the Y$_3$Al$_5$O$_{12}$ garnet phase (hereinafter YAG) (PDF# 000-0033-0040), and the YAlO$_3$ phase (hereinafter YAP) (PDF #000-0033 -0041). The content of the YAP impurity phase does not exceed 5 wt.%. In **Figure 2** YAP particles have a lighter shade.

Amorphization of YAG:Nd ceramics begins under Xe ions irradiation with a fluence of $3\times10^{12}$ cm$^{-2}$ (**Figure 3**); the degree of crystallinity decreases with an increase in the ion fluence. In the diffraction patterns of irradiated ceramics (at fluences of $(3 - 10)\times10^{12}$ cm$^{-2}$), two sets of YAG phase reflections with increased lattice parameters compared to the initial sample are observed.

With increasing fluence, a shift in the position of the diffraction maxima of YAP and strained YAG:Nd relative to the unirradiated state is observed (**Figure 4**). As the fluence increases from $10^{12}$ to $6\times10^{12}$ cm$^{-2}$, the intensity of the XRD peaks decreases. At a fluence of $10^{13}$ cm$^{-2}$, an extended halo from the YAG layer is observed instead of a reflection. This is due to the complete X-ray amorphization of the strained YAG phase, which corresponds to a change in the penetrating ability of the near-surface ceramic layer. A change in the position of the peaks of the studied phases towards smaller angles $2\theta$ is observed, which is associated with an increase in the lattice parameter of this phase from 12.1 to 12.6 Å. A similar picture is observed for the YAP phase, only in this case the position of the YAP peak (121) ceases to change after a fluence of $3\times10^{12}$ cm$^{-2}$.

**Figure 5** shows a graph of the dependence of the relative change in the interplanar distances of YAG (420) and YAP (121) on fluence. This graph can be used to estimate the degree of lattice deformation of YAG and YAP upon irradiation. It can be seen that with an increase in the ion fluence to a value of $10^{13}$ cm$^{-2}$, the deformation of the YAG crystal lattice increases almost linearly, then reaches a constant value. At this moment, active destruction of the crystallinity of the YAG phase in the surface layer begins. For the YAP phase, a nonmonotonic dependence of the degree of microstrain of the crystal lattice on fluence is observed. Note the differences in the intensities of the XRD peak of the initial YAG phase and the peak of the deformed phase (**Figure 5**). The obtained result may indirectly indicate the formation of a gradient defect structure during irradiation, varying from layer to layer.

For a more reliable interpretation of the experimental results, an additional experiment was carried out in symmetric geometry with a longer exposure time and a smaller step $\Delta\theta$. The studies were carried out in the angle range of 30–35° and 65–80° for each sample. **Figure 6** shows a comparison of the XRD reflections of a YAG:Nd sample irradiated with Xe ions with a fluence of $3\times10^{12}$ cm$^{-2}$ at different $2\theta$ angles.

During an experiment in symmetric geometry, the angle of the primary beam and, as a consequence, the information diffraction depth changes continuously. **Figure 6** shows that during the experiment the intensity ratio of the initial and strained phases changes. This may indicate a complex layered structure in the near-surface layer of the sample. **Figure 6** shows the estimated values of the diffraction information depth for different angular ranges (see also [32]). It can be seen that with an information depth of analysis of ~6 μm, the ratio of the intensity of the strained YAG phase to the intensity of the initial phase is ~2:1. At an analysis depth of 12 μm, the intensity ratio of the strained and initial phases approaches 1:1. This suggests that the strained YAG phase is localized in a near-surface layer ~5 μm thick and is formed due to irradiation with Xe ions. The phase of the initial YAG is located deeper in the area that has not received damage from irradiation.

The critical fluence for YAG:Nd ceramics was calculated from the dependence of the relative intensity of the YAG (420) XRD peak on the ion fluence (**Figure 7a**). The calculated value of the critical fluence, determined by linear interpolation of the $I/I_0$–

Fluence dependence in logarithmic coordinates (**Figure 7b**), was $6.5 \times 10^{12}$ cm$^{-2}$. This critical fluence value corresponds to 0.03 dpa (calculated in the SRIM software package [33]).

The microstructure of ceramics after irradiation is shown in **Figure 8**. It can be seen that it does not change; the parameters of the microstructure of irradiated ceramics are similar to the parameters of the microstructure of the initial ceramics (see **Figure 2**). There is no significant increase in the size or volume fraction of pores after irradiation. On the surface of the fractured samples after irradiation with a fluence of $1\times10^{12}$ cm$^{-2}$ (**Figure 8a**) and $3\times10^{12}$ cm$^{-2}$ (**Figure 8b**), a darker layer of partially amorphized material is observed. After Xe irradiation with a fluence of $3\times10^{12}$ cm$^{-2}$, the thickness of the partially amorphized layer reaches ~ 5 μm. When preparing fractured samples after irradiation with a high fluence, the brittle amorphized layer is easily destroyed and breaks off from the surface of the samples (**Figure 8c**).

Microhardness measurements under various loads were carried out to control the nature of changes in the mechanical properties of the surface layer of ceramics after irradiation. By increasing the applied load, the depth of penetration of the indenter into the ceramics surface layers was increased (**Figure 9a**). Thus, at low loads, the microhardness of ceramics characterizes the state of the irradiated surface, and at high loads - the state of the non-irradiated material. From **Figure 9b** it is clear that the microhardness of non-irradiated ceramics practically does not depend on the applied value. After irradiation, the dependence of microhardness on the value of the applied load changes (**Figure 9b**). The microhardness of the surface layer turns out to be less than the microhardness of the central layers of the ceramic sample. This is due to the high density of radiation defects and the presence of areas of the amorphous phase on the surface of the irradiated ceramics. Microhardness $H_V$ of the ceramics surface layer decreases with increasing fluence (**Figure 9b**). It should be noted that the HV of ceramics irradiated with fluences of $1\times10^{12}$ and $3\times10^{12}$ cm$^{-2}$ practically does not differ from the initial one, while the $H_V$ of ceramics irradiated with a fluence of $10^{13}$ cm$^{-2}$ decreases by approximately 1.5 times. Thus, the microhardness measurement technique can be an effective method for monitoring the degree of radiation damage to the surface of ceramics after irradiation.

## 4. Conclusions

Ceramics based on $Y_{2.5}Nd_{0.5}Al_5O_{12}$ oxide were obtained by the SPS method. The resistance of the resulting ceramics to irradiation with Xe ions (E = 146 MeV) in the fluence range $(1-10)\times10^{12}$ ions/cm$^2$ was studied. As a result of irradiation, a gradient defect structure is formed in ceramics, varying from layer to layer. The strained YAG phase formed as a result of Xe ions irradiation is localized in a near-surface layer with a thickness of ~5 μm. The phase of the initial YAG is located deeper, in the area that has not received damage from irradiation.

The calculated value of the fluence leading to complete amorphization was $6.5\times10^{12}$ cm$^{-2}$, which corresponds to 0.03 dpa. The microstructure of ceramics does not change after irradiation. The microhardness of the surface layers of irradiated ceramics is less than the central layers, and, in general, decreases with increasing ion fluence.

**CRedit authorship contribution statement:** Alekseeva L.S.: Investigation (Synthesis), Writing - Original Draft, Writing - Review & Editing, Formal analysis; Nokhrin A.V. & Chuvil'deev V.N.: Data Curation, Writing - Original Draft, Writing - Review & Editing, Formal analysis. Yunin P.A. & Nazarov A.A.: Investigation (XRD);

Orlova A.I.: Methodology, Funding acquisition, Writing - Original Draft, Formal Analysis, Data Curation; Skuratov V.A., Issatov A.T. & Kovylin R.S.: Investigation (Radiation experiments); Murashov A.A.: Investigation (SEM); Boldin M.S. & Voronin A.V. Investigation (SPS); Zotov D.A.: Investigation (Hardness test).

**Conflict of interest**. The authors declare that they have no known conflicts of financial interests or personal relationships that could have appeared to influence the work reported in this paper.

**Funding:** This work was supported by the Russian Science Foundation (grant number 21-13-00308). XRD studies of the samples were carried out in the Laboratory for Diagnostics of Radiation Defects in Solid-State Nanostructures of the Institute of Physics and Mathematics of the Russian Academy of Sciences with the support of the Ministry of Science and Higher Education of the Russian Federation (No. 0030-2021-0030). Work on irradiation of ceramic samples was carried out at JINR (Dubna), with partial support from the Ministry of Science and Higher Education of the Russian Federation (unique project identifier RF-2296.6132X0037).

**Data Availability:** Data will be available on the request.

**List of Figures:**

**Figure 1.** Graphic representation of ceramic sintering diagrams in coordinates "Time t – temperature T – applied pressure P – vacuum pressure Vac".

**Figure 2**. Microstructure of YAG:Nd ceramics: (a) sample surface, (b) sample center.

**Figure 3.** XRD patterns of YAG:Nd ceramics before and after Xe ions irradiation.

**Figure 4.** θ/2θ scan of YAG:Nd ceramics at an angle range of 30-35° with an exposure time of 3 s in one angular position. The signs "*" and "**" indicate XRD reflections (420) of the initial and strained YAG, respectively, the sign "'" indicates XRD reflections (121) of the YAP phase.

**Figure 5.** Relative change in the interplanar distance of the planes (420) and (121) of the YAG and YAP phases, respectively.

**Figure 6.** θ/2θ scan of YAG:Nd ceramics irradiated with Xe ions with a fluence of $3 \times 10^{12}$ cm$^{-2}$, for two angle ranges with an exposure time of 3 s in one angular position (a). Dependence of the X-ray attenuation length on the grazing angle for YAG:Nd ceramics [35] (b).

**Figure 7.** Dependence of the relative intensity of the XRD peak (420) on fluence in ordinary (a) and logarithmic coordinates (b). $I_0$ - intensity of unirradiated sample.

**Figure 8**. Microstructure of YAG:Nd ceramics after Xe ions irradiation with fluence: (a) $1 \times 10^{12}$ cm$^{-2}$; (b) $3 \times 10^{12}$ cm$^{-2}$; (c) $6.5 \times 10^{12}$ cm$^{-2}$; (d) $1 \times 10^{13}$ cm$^{-2}$.

**Figure 9**. Dependence of the calculated indentation depth of the indenter on the load (a) and the dependence of microhardness on the applied load for YAG:Nd ceramics after Xe ions irradiation with different fluences.

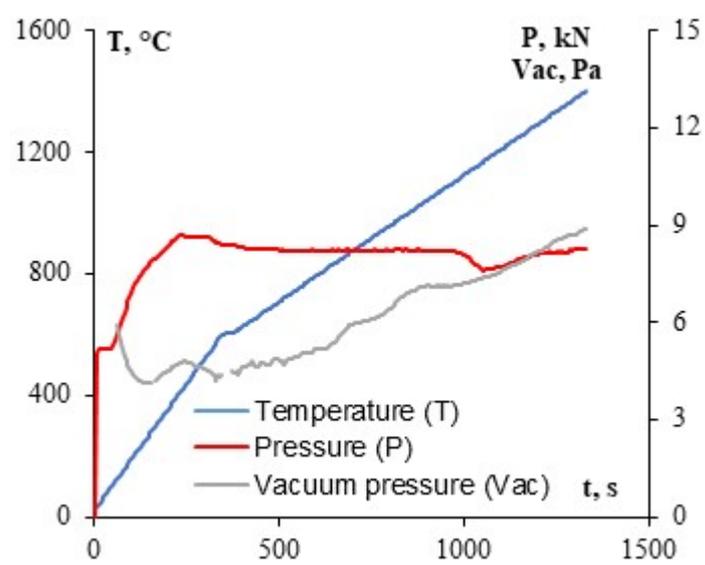

Figure 1

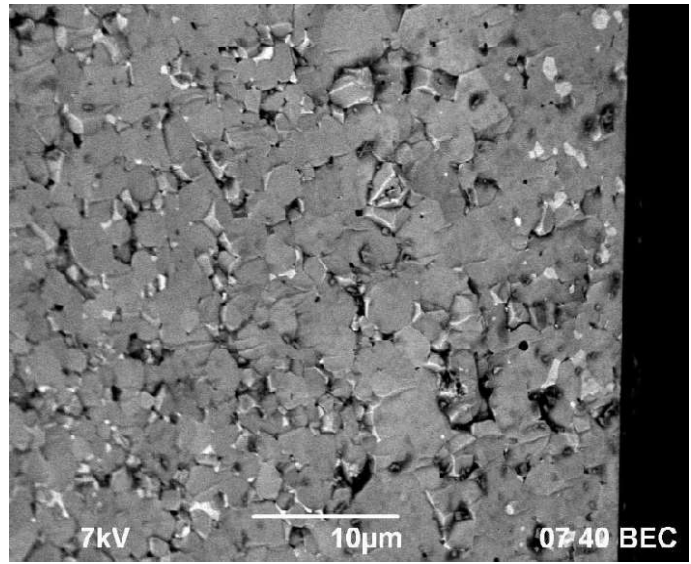
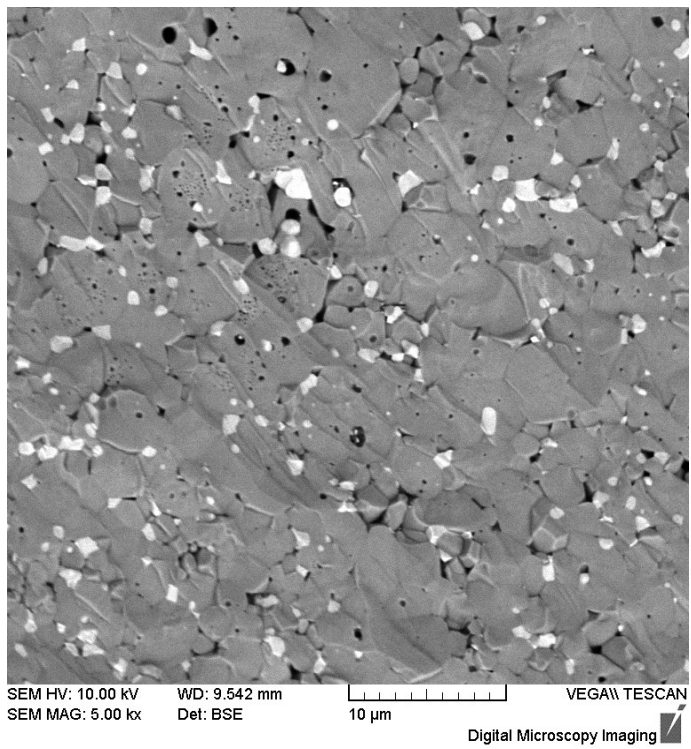

Figure 2

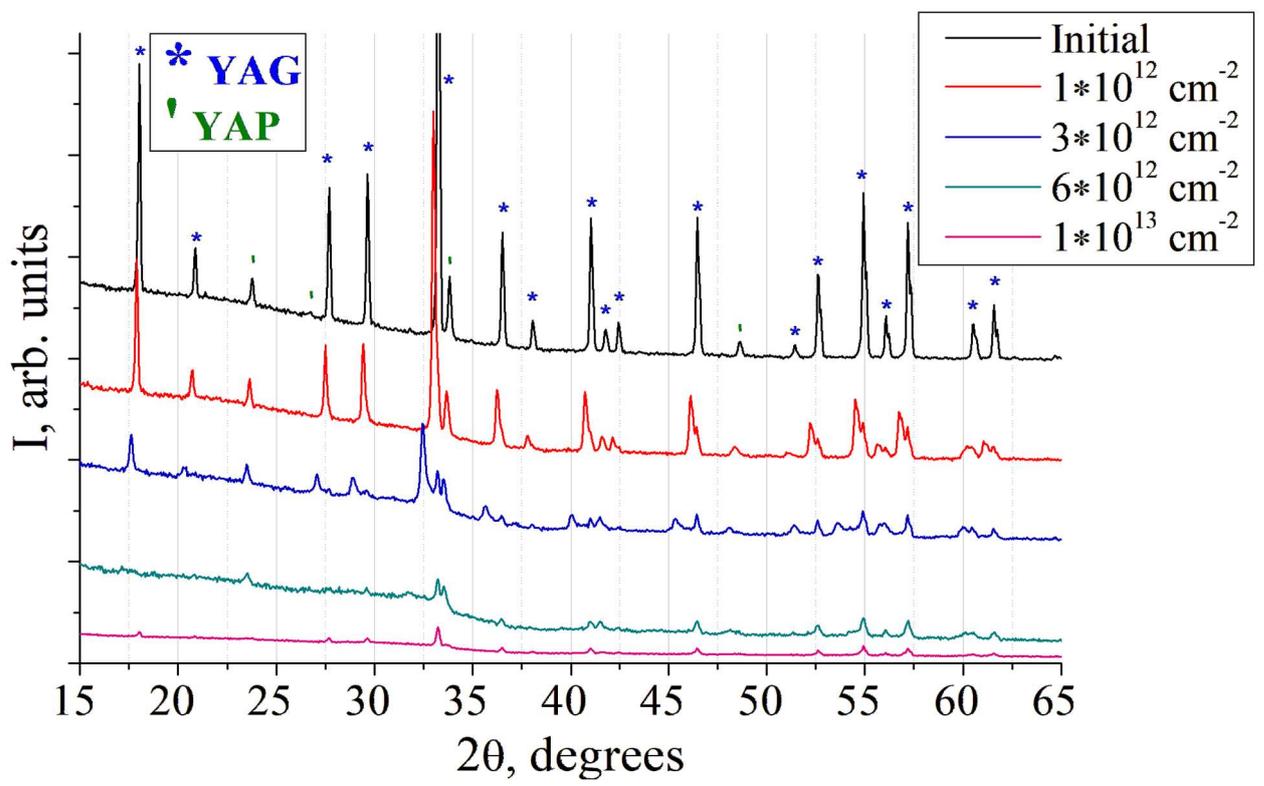

Figure 3

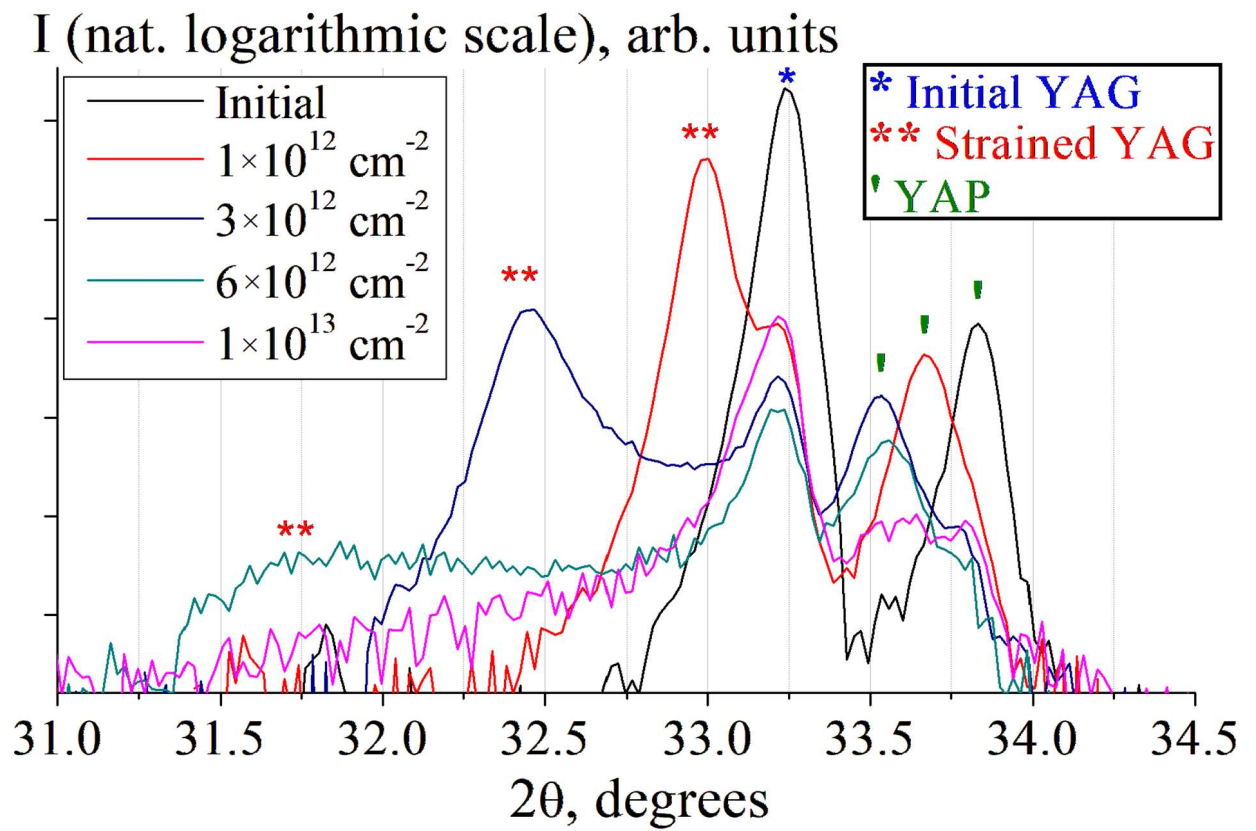

Figure 4

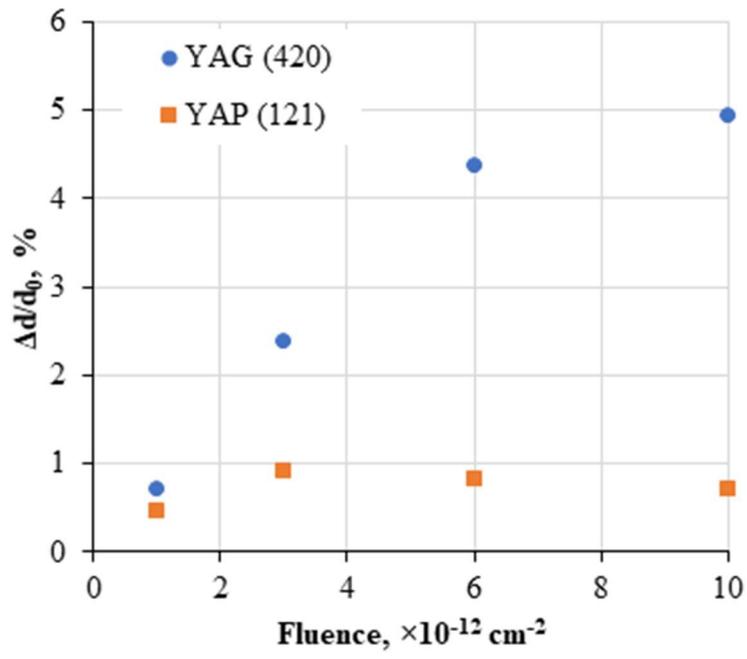

Figure 5

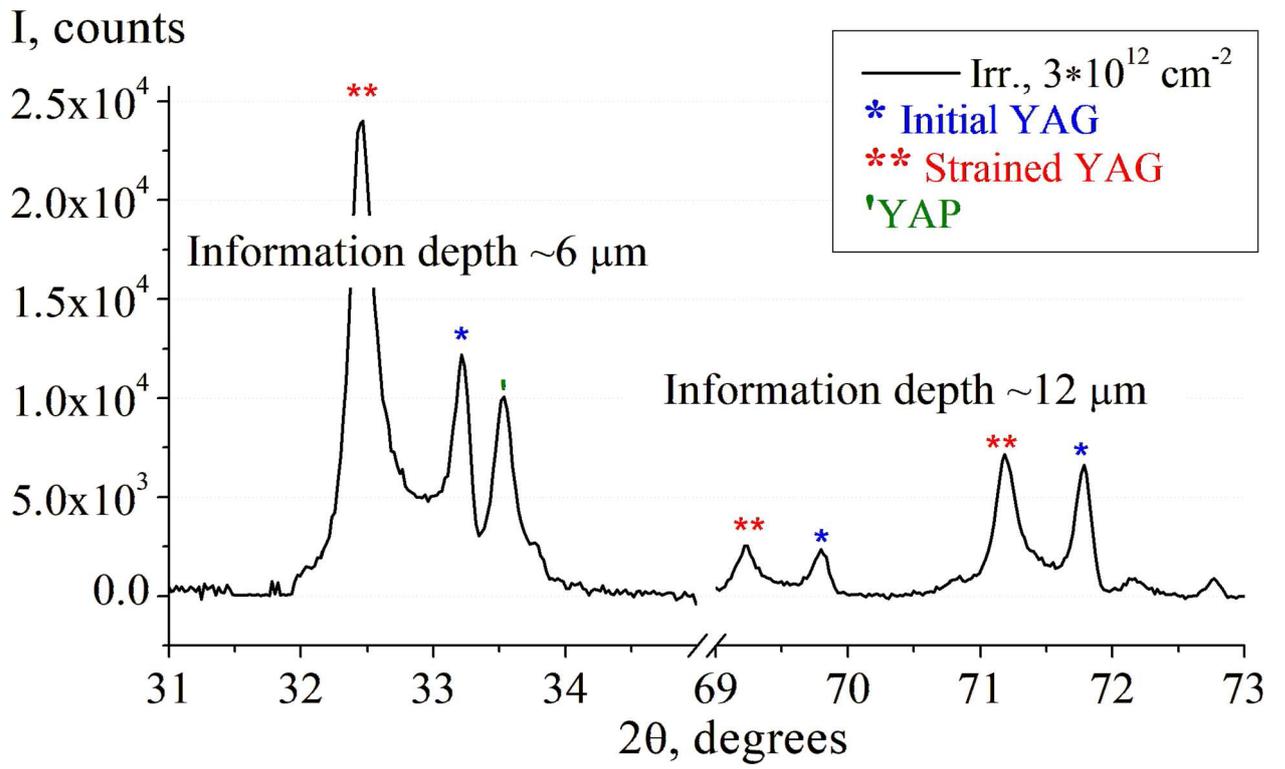

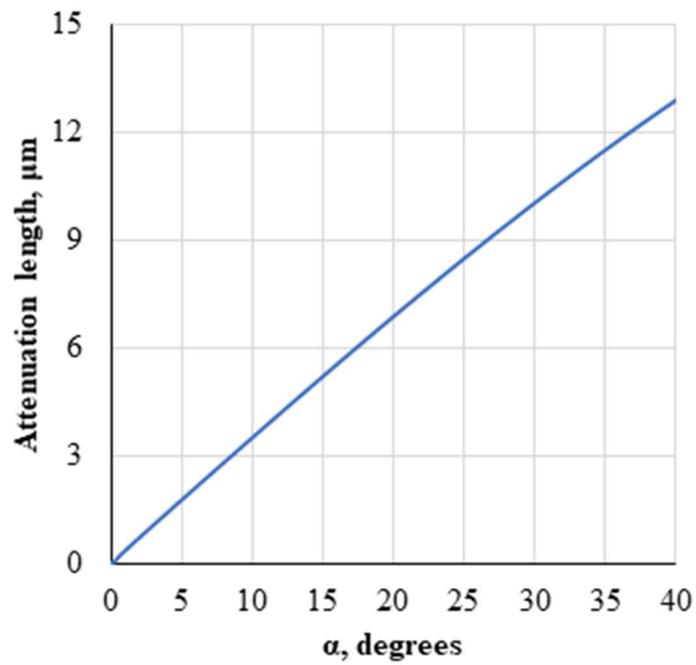

Figure 6

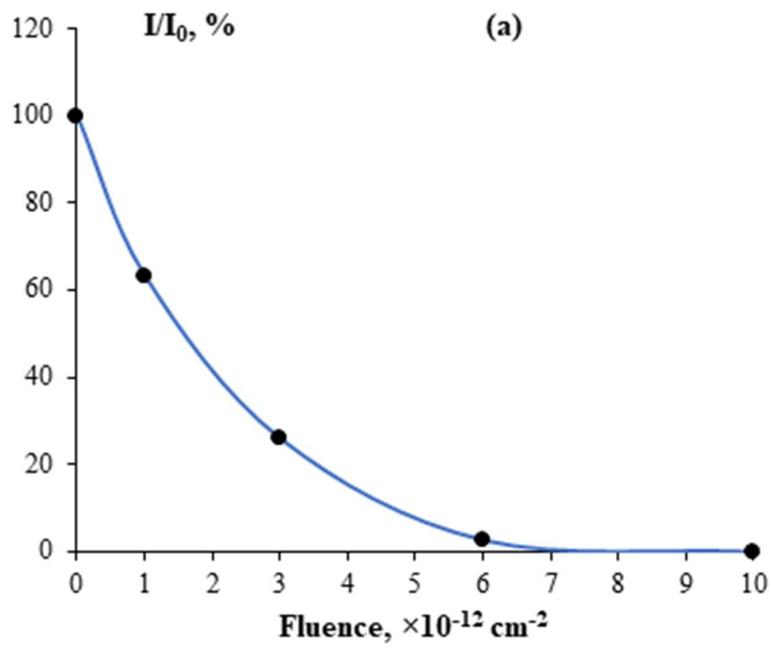

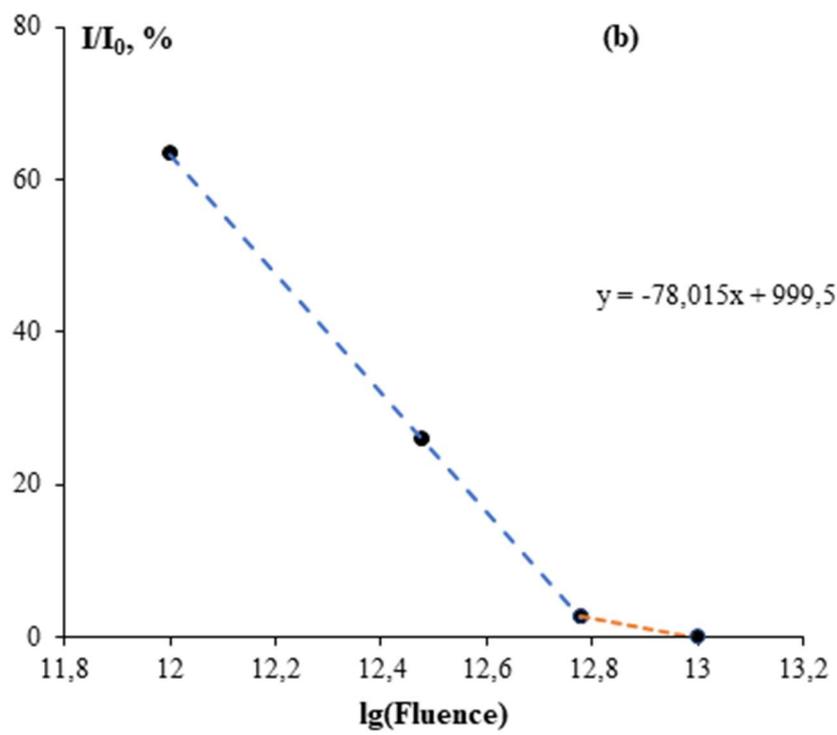

Figure 7

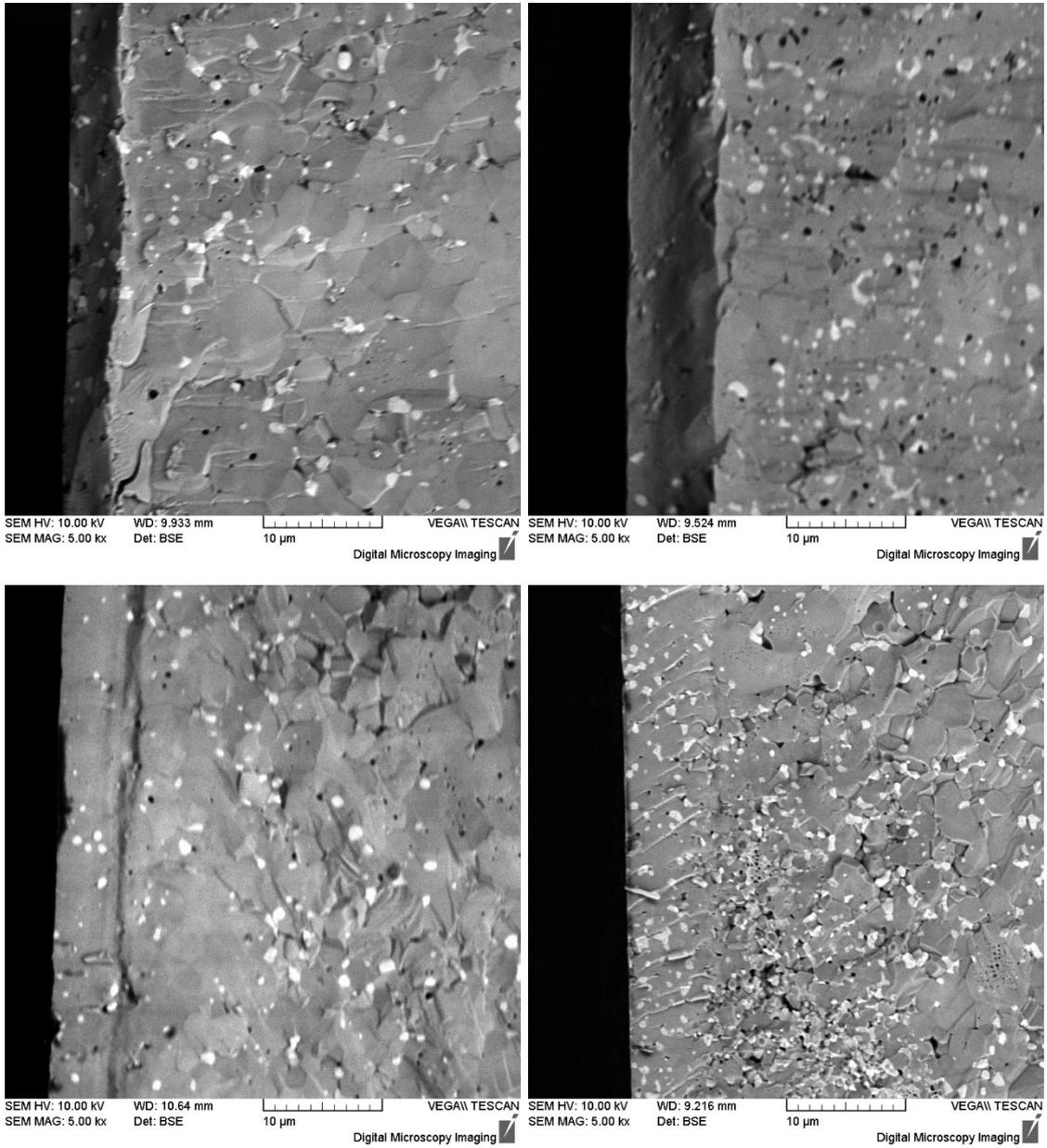

Figure 8

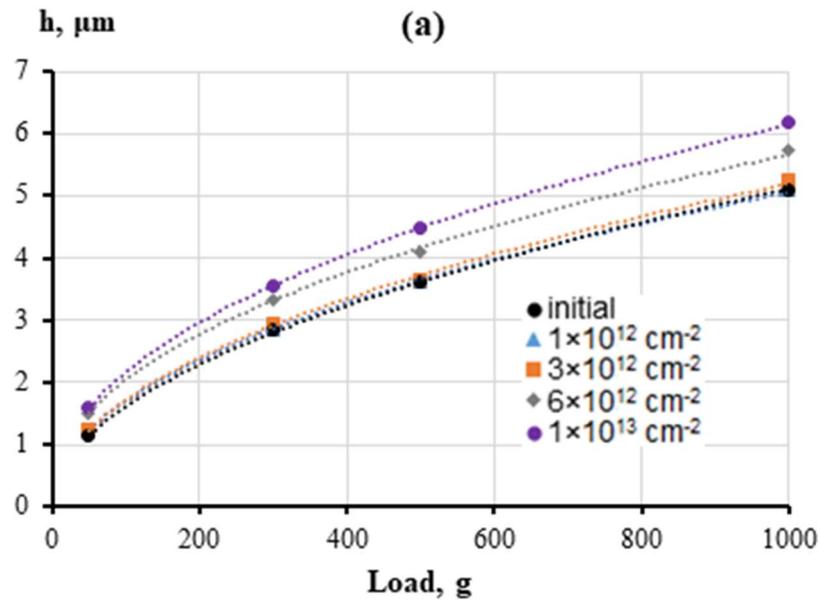
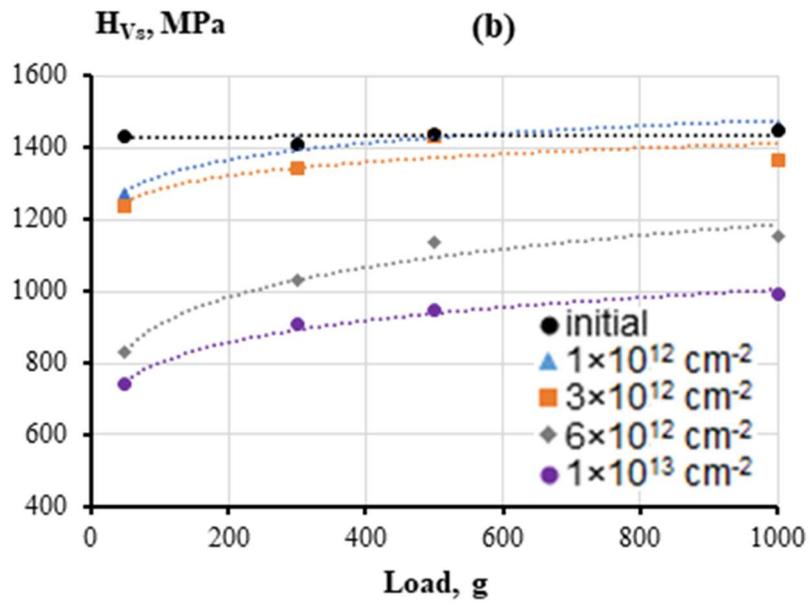

Figure 9